\documentclass[11pt,nofootinbib,aps,superscriptaddress,preprint,floatfix]{revtex4}

\usepackage{graphicx,epsf,amsmath}
\usepackage{bm}

\def\D0bar{\overline D{}^0}
\def\K0bar{\overline K{}^0}

\def\beq{\begin{equation}}
\def\eeq{\end{equation}}
\def\beqa{\begin{eqnarray}}
\def\eeqa{\end{eqnarray}}
\def\bea{\begin{eqnarray}}
\def\eea{\end{eqnarray}}
\def\beq{\begin{equation}}
\def\eeq{\end{equation}}

\def\Re{{\cal R \mskip-4mu \lower.1ex \hbox{\it e}\,}}
\def\Im{{\cal I \mskip-5mu \lower.1ex \hbox{\it m}\,}}
\def\be{\begin{equation}}
\def\ee{\end{equation}}

\def\be{\begin{equation}}
\def\ee{\end{equation}}

\begin{document}

\vspace*{2cm}

\voffset 30pt
\title{\boldmath Unbinding the Deuteron} 

\vspace{1.0cm}

\author{Eugene Golowich}
\affiliation{Department of Physics,
        University of Massachusetts\\[-6pt]
        Amherst, MA 01003}


\begin{abstract}
We consider a description of the deuteron based on 
meson exchange potentials.  A key feature is the inclusion 
of the $I=S=0$ two-pion intermediate state 
(`$\sigma (600)$') as a significant component 
of the inter-nucleon potential energy.  In this approach, deuteron 
binding is seen to be predominantly a consequence of 
$\sigma (600)$ and $\omega (783)$ exchange, with a secondary 
role played by $\rho (770)$.  We explore sensitivity of two-nucleon 
binding to changes in the potential and thereby obtain 
an anthropic constraint ---  that the deuteron unbinds for a modest 
decrease (about $6\%$) in the attractive $\sigma (600)$ 
potential.  
\vskip 1in

\end{abstract}

\def\thepage{{}}
\maketitle
\def\thepage{\arabic{page}}

\section{Introduction}
Imagine a sequence of worlds in which the light quark 
masses are continuously varied away from their physical values. 
It has been argued that heavy nuclei will disassociate for a 
$64\%$ increase in the sum $m_u + m_d$~\cite{dd}.  One can 
anticipate that before this happens, the deuteron will become unbound.  
This is because the deuteron binding energy is less than the average 
binding energy per nucleon.  Here, we study this problem
quantitatively and provide details about how the deuteron would 
respond to variations in its potential induced by changes in  
$m_u + m_d$~\cite{{Beane:2002vs}}.  

The deuteron is an example of a hadronic molecule.  Such a 
two-hadron composite, bound by the strong interactions and 
whose mass lies slightly below the associated two-hadron threshold,  
should be describable with a nonrelativistic 
Schr\"odinger equation.  We can picture the two hadrons as interacting via
a meson exchange potential~\cite{rr}.   Hadronic, as opposed to quark, 
degrees of freedom should be appropriate for these situations.  
In this paper, we consider the $S$-wave radial Schr\"odinger equation, 
\beq\label{schr}
\left[ -{\hbar^2 \over 2 M} {d^2 \over dr^2} + V(r) - E \right] 
u(r) = 0 \ \ , 
\eeq
where $M$ is the reduced mass of the two hadrons 
and $u(0) = u(\infty) = 0$.  

In Section~II, our focus will be 
on the form of the potential energy function $V(r)$, especially as 
regards the contributions from multi-pion exchanges.  
In Section~III, we carry out a numerical study of 
how varying the potential energy affects deuteron binding. 
Our conclusions and plans for future study are presented in Section~IV.

\section{Multi-pion Exchange in Deuteron Binding}\label{sect:deut} 
Although the deuteron state, and more generally the two-nucleon 
potential, are long-studied areas of research, they continue to attract 
theoretical attention.  There has been much recent activity 
about how the nuclear potential behaves in the chiral limit of 
zero light quark mass~\cite{{Epelbaum:2002gb},{Beane:2002xf}, 
{Donoghue:2006rg}}.  This work has motivated our interest
in deuteron binding.  By design our description is simple, 
{\it e.g.} we omit tensor interactions and thus our `deuteron' 
has no D-wave component~\cite{Ericson:1985hf}.  We comment further 
on this aspect of our work in the Conclusion.

Spatial potential energy functions for nonrelativistic calculations
are obtained from particle exchange diagrams in 
quantum field theory.  A nonrelativistic reduction 
of the exchange amplitude is followed by Fourier transformation from 
momentum to position space (see Eqs.~(\ref{vq})-(\ref{vr}) below).  
For example, photon exchange between an electron and a proton leads to 
a number of separate effects (collectively the {\it Breit-Fermi} 
interaction~\cite{dgh}) starting with the long range Coulomb potential.
In addition to generating 
long range potentials, such exchange amplitudes also 
generally give rise to short range or even local contributions.  
In this paper, we restrict our attention to only the long range 
part. 

\subsection{Two-pion and Three-pion Exchange Potentials} 
For convenience, we shall cast the potential energy in a 
dispersive form~\cite{Cottingham:1963}.  
The strongest components of the two-nucleon potential 
arise from the isoscalar ($I=0$) two-pion and three-pion 
exchange channels.  A smaller isovector 
two-pion component must also be included.  

For isoscalar (I=0) potentials, the two-pion 
(scalar) exchange gives rise to an 
attractive interaction while the three-pion 
(vector) interaction is repulsive.  
A momentum space version is given in terms of 
a corresponding spectral function $\rho_i^{\rm (I=0)} (\mu)$,  
\beq
V_i^{\rm (I=0)} (q^2) = {1 \over \pi} \int_{\mu_i^2}^\infty \ d \mu^2 \ 
{\rho_i^{\rm (I=0)} (\mu) \over \mu^2 + q^2} \ \ ,   
\label{vq}
\eeq
where $i = S,V$ and $\mu_S^2 = (2 m_\pi)^2, \mu_V^2 = (3 m_\pi)^2$.
The spatial potential energie $V_i^{\rm (I=0)} (r)$ 
corresponding to Eq.~(\ref{vq}) is 
\beq
V_i^{\rm (I=0)} (r) = {1 \over 4 \pi^2 r} \int_{\mu_i^2}^\infty \ d \mu^2 \ 
e^{-\mu r} \rho_i^{\rm (I=0)} (\mu) \ \ .
\label{vr}
\eeq
The above unsubtracted dispersive forms are taken 
from Refs.~\cite{{Donoghue:2006du},{Donoghue:2006rg}}, where 
behavior at relatively low energies ($< 1$~GeV) is studied. 
The issue of high energy behavior and subtractions is considered 
in Ref.~\cite{Adams:2008hp}.  However, the choice of a subtraction 
scale at sufficiently high energy is not expected to alter
the basic conclusions of 
Refs.~\cite{{Donoghue:2006du},{Donoghue:2006rg}}, which are anchored 
by the constraints of chiral perturbation theory~\cite{jfd0}.

For the isoscalar vector channel, it is a good approximation to take 
$\rho_V^{\rm (I=0)} (\mu) \propto 
\delta(\mu - m_\omega)$, where $m_\omega \simeq 783$~MeV.  
This is also appropriate for the isovector two-pion exchange 
channel, where now $\rho_V^{\rm (I=1)} (\mu) \propto 
\delta(\mu - m_\rho)$, with $m_\rho \simeq 770$~MeV.  
In these two cases, the association between an exchanged vector meson of 
mass $m_V$ and a spatial Yukawa potential $V_V(r) \propto e^{-m_V r}$ 
is natural and obvious.  The isoscalar two-pion exchange channel 
is more subtle.

\subsection{The Two-pion Isoscalar (`$\bm{\sigma} (600)$') 
Component of the Two-nucleon Potential} 
The nucleon-nucleon potential is known to have 
a strong attractive contribution of intermediate range.  This  
arises from an exchange potential in the scalar, isoscalar 
channel, due to the exchange of a pair of pions in the 
$I = S = 0$ channel.  There is now a description of this which, 
although not explicitly given in terms of quark interactions, 
does nonetheless respect the underlying QCD dynamics. 
The approach is (i) to use chiral perturbation theory 
to describe the low energy part of the momentum-space potential, 
and (ii) to invoke two-pion rescattering via the Omnes function 
at higher energies in order to include the constraint of unitarity. 

The chiral component of $\rho_S^{\rm (I=0)}$ turns out, as is 
generally the case in chiral perturbation theory, to 
depend on a small number of low energy constants whose 
numerical values are obtained phenomenologically.  
To the extent that these constants depend on any meson, it is 
actually $\rho(770)$ that contributes.  The chiral form 
of $\rho_S^{\rm (I=0)}$ grows 
with energy and at some point requires modification in order to 
respect unitarity.  Such a modification can be 
accomplished~\cite{Donoghue:2006rg} by introducing the Omnes function, 
\beqa
& & \Omega^{\rm (I=0)} (\mu) = 
\exp \left[ {\mu^2 \over \pi} \int {ds \over s}
~{\delta^{\rm (I=0)} (s) \over s - \mu^2}\right] \ \ ,
\label{omnes} 
\eeqa
where $\delta^{\rm (I=0)} (s)$ is the S-wave two-pion phase shift in the 
$I=0$ channel.~\footnote{It is also possible to proceed 
to a higher order of the chiral expansion~\cite{Colangelo:2001df}.}    
In this manner, the two ingredients of chiral perturbation theory 
and two-pion rescattering provide a sound theoretical basis for 
understanding the two-pion $S=I=0$ exchange potential.  

In the intermediate energy range $400 \le E_{\pi\pi} \le 700$~MeV, 
the $S=I=0$ two-pion phase shift does not pass through 
$90^{\rm o}$ and thus (unlike $\omega (783)$ and $\rho(770)$)
has no obvious association with a resonant state.    
The dispersive spatial potential of Eq.~(\ref{vr}) 
would seem to require a broad spectral function in this case. 
However, it can be shown (see especially 
Fig.~3(b) of Ref.~\cite{Donoghue:2006rg}) that 
{\it as a numerical recipe} the effect of nonresonant $S=I=0$ two-pion 
scattering is reproduced by 
using $V_\sigma(r) \propto e^{-m_\sigma r}$ with $m_\sigma \simeq 600$~MeV.
In our work, we adopt this approach and refer to it hereafter as 
the `$\sigma (600)$ exchange potential'.  We stress that our use of 
`$\sigma (600)$' is simply a convenient shorthand and not meant to 
commit us to any specific model of this state~\cite{Ding:2004ty}.

\section{Binding the Deuteron}
Let us now consider the two-nucleon interaction in terms of four 
exchange potentials, 
\beq\label{deutpotl}
V^{\rm (NN)} = V_{\sigma}^{\rm (NN)} + V_{\rho}^{\rm (NN)} 
+ V_{\omega}^{\rm (NN)} + V_{\pi}^{\rm (NN)} \ \ .
\eeq
Each of these is taken to have the Yukawa form, 
\beq\label{yuk}
V^{\rm (NN)}(r) = \sum_i\ \eta_i {g_i^2\over 4 \pi} ~ 
{e^{- m_i r} \over r}  \ \ , 
\eeq
where $\eta_i = +1$ for repulsion and $\eta_i = -1$ for attraction.   
All that remains is to fix the strength of each potential.  

For detailed discussions of potential energy fits to nuclear binding, 
see Refs.~\cite{Donoghue:2006du,fs}.  We will assume the following 
values for the $\sigma(600)$ and $\omega(783)$ couplings~\cite{jfd},  
\beq\label{sig-omeg}
{g_{\omega {\rm NN}}^2 \over 4 \pi} \simeq 16.6 \qquad {\rm and} \qquad
{g_{\sigma {\rm NN}}^2 \over 4 \pi} \simeq 10.3 \to 12.9 \ \ .
\eeq

Our strategy will be to fit $g_{\sigma {\rm NN}}^2/ 4 \pi$ to the 
experimental deuteron binding energy~\cite{PDG}  
\beq\label{BEd}
{\rm BE}_{\rm deut} = 2.22457~{\rm MeV} \quad 
\text{(with negligible error)}  
\eeq
and see how it compares to the range in Eq.~(\ref{sig-omeg}).  
We use this approach because a precise fit will allow us to study 
the effect on ${\rm BE}_{\rm deut}$ of modifying the potential 
energy function.  

The pion potential energy is obtained by starting from 
the field theory interaction with pseudoscalar coupling
\beq\label{ps}
{\cal L}_{\pi {\rm NN}} = i f_{\pi {\rm NN}} 
{\bar N} \gamma_5 \bm{\tau}\cdot\bm{\pi} N
\eeq
and reducing the pion exchange graph to its nonrelativistic limit.  
Upon keeping only the Yukawa dependence of Eq.~(\ref{yuk}), one
obtains the isospin-dependent, spin-dependent potential energy
\beq\label{relate}
V_\pi^{\rm (NN)} (r) = {f^2_{\pi {\rm NN}} \over 4 \pi} 
\left( {m_\pi \over 2 M_N} 
\right)^2 (2 {\bf I}^2 - 3) \cdot {2 {\bf S}^2 - 3 \over 3} ~
{e^{- m_\pi r} \over r} 
\ \ .
\eeq 
For the deuteron channel ($I = 0$, $S = 1$) this reduces to 
\beq\label{gpi}
V_\pi(r)^{\rm (NN)} 
= - {g_{\pi{\rm NN}}^2\over 4 \pi} ~ {e^{- m_\pi r} \over r}  \ \ , 
\eeq
with 
\beq\label{pioncoup}
g_{\pi{\rm NN}}^2 \equiv f^2_{\pi {\rm NN}} 
\left( {m_\pi\over 2 M_N} \right)^2 \simeq 0.073 \ \ .
\eeq
Observe that $g_{\pi {\rm NN}}$ is tiny relative to 
$g_{\sigma{\rm NN}}$ and $g_{\omega{\rm NN}}$.

For the $\rho(770)$ interaction the lagrangian is 
\beq\label{vnn}
{\cal L}_{\rm \rho NN} = 
g_{\rho{\rm NN}} {\bar N} \gamma_\mu \bm{\tau}\cdot\bm{\rho}^\mu N \ \ .
\eeq
where we take $g^2_\rho/4\pi = 2.0$~\cite{Jido:2002ig}.    
The $\rho (770)$ component of the potential energy is isospin 
dependent, 
\beq\label{vrho}
V_\rho^{\rm (NN)} (r) 
= {g^2_{\rho{\rm NN}} \over 4 \pi} (2 {\bf I}^2 - 3) ~
{e^{- m_\rho r} \over r} \ \ .
\eeq

\subsubsection{\bf{The S=1, I=0 (Deuteron) Channel}}
\begin{figure}[htbp]
\includegraphics[width=20pc]{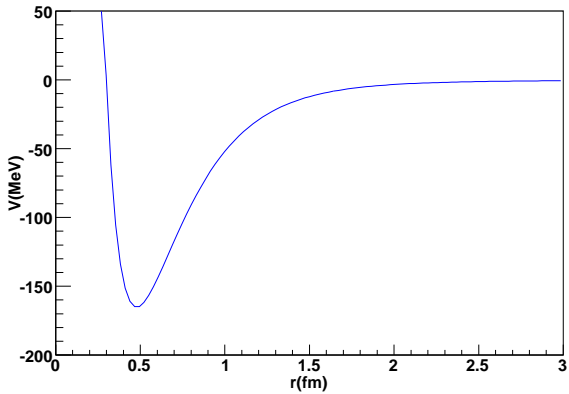}
\caption{Deuteron Potential Energy.}
\label{fig:deut-b}
\end{figure}

Of the four component potentials, only the $\omega (783)$ piece 
is repulsive in the $I=0$ channel.  It is mainly the interplay 
of the large $\omega (783)$ 
repulsion and $\sigma (600)$ attraction that results in the 
deuteron binding.  Although the coefficient of the $\omega (783)$ 
potential is larger than that of the attractive $\sigma (600)$, 
the lighter mass of the $\sigma (600)$ allows it to dominate over 
$\omega (783)$ at intermediate distance scales.  
The $\omega (783)$ provides the repulsive core.  The potential 
energy which includes all four components is displayed in 
Fig.~\ref{fig:deut-b}.  When we use as input potential energies 
$V_\sigma^{\rm (NN)}$, $V_{\omega}^{\rm (NN)}$, 
$V_{\rho}^{\rm (NN)}$ 
and $V_{\pi}^{\rm (NN)}$ to fit the value of 
$g_{\sigma{\rm NN}}^2 /4 \pi$ to the 
deuteron binding energy, the value obtained for 
$g_{\sigma{\rm NN}}$ is 
\beq\label{sigma}
{g_{\sigma{\rm NN}}^2 \over 4 \pi} \simeq 10.83 \ \ .
\eeq
This is in accord with the range given above in Eq.~(\ref{sig-omeg}).

\subsubsection{\bf The S=0, I=1 Channel}
The two-nucleon system has isospins 
$I = 0,1$ and spins $S = 0,1$, and if the two nucleons are in an 
S-wave, the $I = S = 0$ and $I = S = 1$ 
configurations are ruled out by Fermi-Dirac statistics.  
The deuteron has $I=0$, $S=1$, which leaves finally 
the combination $I=1, S=0$.   The $I=1, S=0$ 
channel has a virtual bound state, as evidenced by 
the large scattering length $a(^1S_0) \simeq 23.7$~fm.  
Since $\sigma (600)$ and $\omega (783)$ are isoscalars, a 
potential energy with only these exchanges would imply equal 
binding in both the $I=0, S=1$ and $I=1, S=0$ channels.  However, the 
$\rho (770)$ contribution is attractive in the deuteron channel 
but repulsive (at one third the strength) in the 
$I=1, S=0$ channel.  Our numerical study obtains the desired 
result of no binding for the $I=1, S=0$ channel upon using the 
$\sigma (600)$ coupling strength of Eq.~(\ref{sigma}).

\subsection{Anthropic Implications}
Anthropics can be viewed as relating to the class of theories 
(including possibly string theory and chaotic inflation) in 
which spacetime has a domain structure, each domain with its own 
set of fundamental parameters (CKM mixing angles, lepton and 
quark masses, {\it etc})~\cite{ab}.  Such a description is often referred 
to as the {\it Multiverse}~\cite{sw2,jfd2}.  This theoretical possibility 
provides an interesting framework for viewing deep issues 
of physics, such as the cosmological constant~\cite{Weinberg:1987dv}, 
the Higgs vacuum expectation value~\cite{abds} and 
the light quark masses~\cite{dd}.  

Previous studies have revealed that even a modest variation 
of fundamental parameters can induce qualitative changes in 
physical systems~\cite{dd,abds}.  Works like these 
can help determine the `physical' windows in parameter space 
which enable Universes like our own to exist.  This is in contrast
with `theoretical' parameter windows which can be associated 
with a given underlying theory, {\it e.g.} string dynamics.  If the 
physical window is small compared to the theoretical window, then 
it is not unreasonable to assign a flat probability distribution 
across the physical window and thus to study the relative liklihood of 
that our Universe should exist within a given theoretical 
framework~\cite{sw2}.  

\begin{table}
\centering
\begin{tabular}{|c|c||c|c|} \hline\hline
$g_{\sigma NN}^2/(4\pi)$ 
& BE$_{\rm deut}$ (MeV) & $R^{\rm (coup)}$ & 
$R^{\rm (BE)}$ 
\\ \hline
$10.83$ & $2.226$ & $1.$ & $1.$\\
$10.81$ & $2.032$ & $0.998$ & $0.913$ \\ 
$10.71$ & $1.422$ & $0.989$ & $0.639$ \\ 
$10.63$ & $1.019$ & $0.982$ & $0.457$ \\ 
$10.59$ & $0.813$ & $0.977$ & $0.365$ \\ 
$10.53$ & $0.611$ & $0.972$ & $0.274$ \\ 
$10.47$ & $0.408$ & $0.967$ & $0.183$ \\ 
$10.38$ & $0.206$ & $0.959$ & $0.092$ \\ 
$10.32$ & $0.105$ & $0.953$ & $0.047$ \\ 
$10.21$ & $0.012$ & $0.943$ & $0.006$ \\ \hline\hline
\end{tabular}
\caption{\label{tab:rstab}
Dependence of deuteron binding energy on the $\sigma NN$ coupling as 
expressed directly (first two columns) or in terms of the 
ratios $R^{\rm (coup)}$, $R^{\rm (BE)}$ defined in Eq.~(\ref{ratio}).}
\end{table}

In this paper, we are concerned with the light quark masses.  
Refs.~\cite{jfd,dd} show that increasing $m_u + m_d$ decreases the 
$\sigma NN$ coupling whereas decreasing $m_u + m_d$ has the opposite 
effect.  For example, it is estimated that taking 
$m_\pi^2 \to 0$ produces a $40\%$ increase in 
$g_{\sigma{\rm NN}}^2 /4 \pi$ while increasing $m_\pi^2$ to twice its 
physical value decreases $g_{\sigma{\rm NN}}^2 /4 \pi$ by $20\%$.  
The source of this sensitivity to $m_\pi$ (and ultimately to $m_u + m_d$) 
lies mainly in the threshold contribution 
in the dispersive integrals of Eqs.~(\ref{vq}),(\ref{vr}).  
By contrast, the $\rho$NN and $\omega$NN couplings are relatively 
unaffected by such changes in the light quark masses because 
their spectral functions are large only well above threshold.   
\begin{enumerate}
\item {\it Unbinding of the Deuteron via Weakening of 
$V_\sigma^{\rm (NN)}$}: 
In our description, the two-nucleon potential is mainly the
competition between the attractive $V_\sigma$ and the repulsive $V_\omega$.  
It is not hard to upset this balance and thus unbind the deuteron 
by varying some combination of the 
potential energies, {\it e.g.} by reducing 
$V_\sigma^{\rm (NN)}$ (keeping all other interactions fixed),  
increasing $V_\omega^{\rm (NN)}$ 
(again keeping all other interactions fixed),
{\it etc}.  However, the former is the only variation associated 
with the mass values of the light quarks.
To help keep track of how the deuteron binding energy responds to 
the variation in $g_{\sigma NN}$, we find it useful 
to define the following ratios, 
\beq\label{ratio}
R^{\rm (coup)}[g_{\sigma NN}] \equiv 
{g_{\sigma NN}^2/(4\pi) \over 10.83} 
\qquad 
R^{\rm (BE)}[BE_{\rm deut}] \equiv 
{{\rm BE}_{\rm deut}~{\rm (MeV)} \over 2.263} \ \ .
\eeq
From the numerical values in 
Table~\ref{tab:rstab}, we see that the deuteron becomes unbound  
if $V_\sigma^{\rm (NN)}$ is reduced in strength by about $6\%$.  This is 
to be compared with the $10\%$ required 
to disassociate heavy nuclei~\cite{dd}.  

\item {\it Alternative Variations}: 
There can, in principle, be other parameter variations.
For the sake of completeness, we briefly consider a few of these.  
\subitem It is obvious that increasing 
$V_\sigma^{\rm (NN)}$ will increase the 
deuteron binding energy.  However, there is another consequence,  
which turns out to somewhat alter the physical world.  
We find that an increase of about $8 \%$ in 
$V_\sigma^{\rm (NN)}$, with other potentials 
held fixed, suffices to first produce binding the $I=1, S=0$ 
NN channel. 
\subitem Varying the $\omega$ and $\rho$ potentials would likewise have 
physical consequences, {\it e.g.}  with $V_\sigma^{\rm (NN)}$ 
fixed, an increase in strength of $V_\omega^{\rm (NN)}$ 
by about $7\%$ would result in unbinding the deuteron, 
This kind of variation is not associated 
in any obvious way with the fundamental parameters of the 
Standard Model and thus has no connections with our anthropic analysis.  
\end{enumerate} 

\section{Conclusion}

Our aim in this paper has been to explore the dependence of 
deuteron binding on variations in $m_u + m_d$ in terms of a 
simple model involving just 
central potentials.  The central potentials were constructed 
with care, by exploiting up-to-date results from studies of nuclear 
binding.  This description is admittedly in contrast with the traditional
approach which uses the one-pion-exchange (OPE) tensor 
potential~\cite{Ericson:1985hf}.  
We feel that these two procedures are not incompatible. 
Consider a description containing the huge $\sigma$ and 
$\omega$ potentials along with the tensor OPE.  
If the $\sigma$ and $\omega$ effects cancel even more than in our 
calculation (which would involve a slight change in $V_\sigma$), 
the deuteron binding would be given largely in terms of the tensor OPE.  
Even so, deuteron binding would still feel the changes induced by  
$m_u + m_d$ largely via the effect on $V_\sigma$ (as described in 
this paper).  We leave for future study a more ambitious description 
which includes all these elements in a Schr\"odinger equation context, 
plus insights from chiral perturbation theory.

Thus, given the deuteron potential energies of Eq.~(\ref{deutpotl}), 
we have shown ({\it cf} Table~\ref{tab:rstab}) that a slight 
reduction in $V_\sigma$ results in unbinding the deuteron.  
Even a slight weakening of the attractive component results in 
a large effect.  As a result, the deuteron is a sensitive 
system for studying how physics would respond to changes in 
$m_u + m_d$.  The crucial links in the chain of logic which connects these 
are (i) the chiral relation, $m_\pi^2 = B_0 (m_u + m_d)$, which relates 
the pion mass to the light quark masses and (ii) 
the dispersive formula Eq.~(\ref{vq}) which relates 
$V_\sigma$ to the pion mass~\cite{Donoghue:2006du}.
The $6\%$ decrease in $V_\sigma$ needed to unbind deuterons 
is less than the estimated $10\%$ needed to unbind heavy 
nuclei~\cite{dd}.

It is abundantly clear that if conditions forbade atoms or 
nuclei from existing, then there would be no Universe as we know 
it. The absence of the deuteron, although itself just a solitary 
quantum particle, would likewise be dramatic.  The deuteron first appears 
in Cosmology as part of Big Bang Nucleosynthesis (BBN) in the early 
Universe, 
\beq\label{bbn0}
p + n \ \rightleftharpoons \  D + \gamma \ \ .
\eeq
Before deuteron formation, free neutrons are copious (about one
neutron per five protons), but after formation free neutrons are 
absent (the number of free neutrons equals the number of deuterons at 
about 200 seconds after the Big Bang).  Once deuterons are formed, 
they take part in a number of additional BBN reactions, such as 
fusion with protons and neutrons, 
\beq\label{bbn1}
D + p \ \rightleftharpoons \ ^3{\rm He} + \gamma \qquad 
D + n \ \rightleftharpoons \ ^3{\rm H} + \gamma  \ \ , 
\eeq
or with each other, 
\beq\label{bbn2}
D + D \ \rightleftharpoons \ ^3{\rm H} + p \qquad 
D + D \ \rightleftharpoons \ ^3{\rm He} + n \ \ .
\eeq
These lead to the production of $^4{\rm He}$, 
\beq\label{bbn3}
\begin{array}{l}
 ^3{\rm H} + p  \ \rightleftharpoons \ ^4{\rm He} + \gamma \\
 ^3{\rm H} + D  \ \rightleftharpoons \ ^4{\rm He} + n 
\end{array} 
\qquad\qquad 
\begin{array}{l}
 ^3{\rm He} + n  \ \rightleftharpoons \ ^4{\rm He} + \gamma \\
 ^3{\rm He} + D  \ \rightleftharpoons \ ^4{\rm He} + p 
\end{array} 
\eeq
Although trace amounts of heavier nuclei are formed, the above 
relations display the basic yield of BBN once the deuteron is 
formed.  It is clear that, although just a 'solitary' particle, 
the deuteron occupies a central role in BBN.\footnote{We restrict 
our comments to BBN, but the deuteron plays a significant 
role in solar fusion.}

There are some additional issues, not yet mentioned, 
which we leave for further study. 
As regards BBN, it would be interesting to study more carefully 
what happens under conditions where the deuteron becomes 
slightly unbound.  In a different direction, by 
using a modification of the meson exchange approach described 
in this paper, we can 
consider the possibility of molecular structures in the meson
sector.  Of particular interest are the 
states $a_0(980)$ and $f_0(980)$, whose nearness to the 
$K{\bar K}$ threshold has suggested a molecular 
interpretation~\cite{Weinstein:1990gu}.  We will report 
on this study elsewhere~\cite{eg08}.

\acknowledgments

This work was supported in part by the U.S.\ National Science
Foundation under Grant PHY--0555304.  We thank G. Colon and 
E. Thompson for help with the figure.  Discussions with John 
Donoghue were useful, informative, and above all, interesting.
The author thanks Daniel Phillips for his critical review of the 
manuscript and various insightful remarks.

\end{document}